\documentclass{article}
\usepackage[preprint]{spconf,amsmath,graphicx}
\usepackage{booktabs}
\usepackage{amssymb}
\usepackage{pifont}
\newcommand{\cmark}{\ding{51}}%
\newcommand{\xmark}{\ding{55}}%
\usepackage[hidelinks]{hyperref}
\usepackage{siunitx}
\usepackage{multirow}

\title{Real-time Low-latency Music Source Separation using Hybrid Spectrogram-TasNet}
\name{Satvik Venkatesh, Arthur Benilov, Philip Coleman, Frederic Roskam}
\address{L-Acoustics, 67 Southwood Lane, London N65EG.}
\copyrightnotice{\begin{tabular}[t]{@{}l@{}}© 2024 IEEE. Personal use of this material is permitted. Permission from IEEE must be obtained for all other uses, in any current or\\future media, including reprinting/republishing this material for advertising or promotional purposes, creating new collective works,\\for resale or redistribution to servers or lists, or reuse of any copyrighted component of this work in other works.\end{tabular}}

\begin{document}
\ninept
\maketitle
\begin{abstract}
There have been significant advances in deep learning for music demixing in recent years. However, there has been little attention given to how these neural networks can be adapted for real-time low-latency applications, which could be helpful for hearing aids, remixing audio streams and live shows. In this paper, we investigate the various challenges involved in adapting current demixing models in the literature for this use case. Subsequently, inspired by the Hybrid Demucs architecture, we propose the Hybrid Spectrogram Time-domain Audio Separation Network (\mbox{HS-TasNet}), which utilises the advantages of spectral and waveform domains. For a latency of 23~ms, the HS-TasNet obtains an overall signal-to-distortion ratio (SDR) of 4.65 on the MusDB test set, and increases to 5.55 with additional training data. These results demonstrate the potential of efficient demixing for real-time low-latency music applications.

\end{abstract}
\section{Introduction}
\label{sec:intro}
Audio source separation is the process of separating a mixture of audio into its individual components. This is useful for various applications such as speech enhancement~\cite{defossez2020real}, speech separation~\cite{luo2018tasnet}, dereverberation~\cite{luo2018real}, and as a pre-processing step for sound event detection~\cite{turpault2021sound}. Music source separation (MSS), also popularly known as demixing or unmixing, enables us to remix the balance within the music, perhaps to make the vocals louder or to suppress an unwanted sound, change the spatial location of a musical instrument~\cite{cano2018musical}, create loops from individual instruments~\cite{smith2018nonnegative}, object-based mixing~\cite{coleman2018audio} or even upmix a 2-channel stereo recording to a multi-channel surround sound system. In recent years, there has been tremendous interest in using deep learning for MSS, leading to significant improvements in separation quality. These studies have modelled MSS as a supervised learning problem, where the constituting components are known beforehand. For example, \emph{vocals}, \emph{drums}, \emph{bass}, and \emph{other}. However, the computational cost and latency of these models are generally high~\cite{hennequin2020spleeter,defossez2019music}, making them suitable for offline tasks and studio productions, but not for real-time low-latency scenarios. 

Before the onset of deep learning, \cite{barry2004real} presented Azimuth Discrimination and Resynthesis for real-time MSS. The window and hop sizes for the short-time Fourier transform (STFT) were set to 93~ms and 23~ms respectively. Therefore, considering the overlap-add operation of the inverse-STFT, the latency of the algorithm is 93~ms. In \cite{tahmasebi2020design}, a multi-layer perceptron (MLP) with an algorithmic latency of 23~ms performed singing voice separation. Their goal was to remix music for cochlear implant users. Unlike MSS, researchers have widely explored deep learning for real-time speech separation~\cite{luo2018tasnet} and enhancement~\cite{defossez2020real}. For STFT-based speech enhancement algorithms, typically a latency of 32~ms is observed, due to the required window size~\cite{luo2019conv, wang2022stft}. End-to-end deep learning directly on raw audio for speech separation has obtained impressive results with latency as low as 2~ms~\cite{luo2019conv}. However, low-latency MSS is more challenging due to higher sampling rates, high dependence on future context, and potentially larger models. 

Neural network architectures for MSS can be broadly classified into spectrogram-based, waveform-based, and hybrid approaches. Initially, spectrogram-based models only used the magnitude and discarded the phase~\cite{jansson2017singing, stoller2018wave, sawata2021all}. The network predicted the magnitude mask and used the phase of the input mixture for the final output. The waveform-based models that directly separate raw audio have the advantage of not discarding phase and extracting potentially useful deep features~\cite{stoller2018wave, defossez2019music}. 
Furthermore, studies have explored ways to predict phase~\cite{kong2021decoupling} and also predict real and imaginary parts of the spectrogram~\cite{luo2022music} to obtain state-of-the-art results. Recently, studies have also proposed hybrid frameworks that harness the advantages of both waveform and spectral domains~\cite{kim2021kuielab, defossez2021hybrid}.

In this paper, we investigate real-time MSS with a low-latency such as 23~ms (frame size of 1024 samples at 44.1kHz). To our knowledge, this is the first paper that explores deep learning to separate \emph{vocals}, \emph{drums}, \emph{bass}, and \emph{other} in real-time with low-latency.
Using 23~ms was a good starting point because it is large enough for the Fourier spectrum to capture meaningful spectral features, but small enough for real-time applications such as cochlear implants~\cite{tahmasebi2020design} and stereo upmixing~\cite{barry2004real,barry2019real}. First, we adapt the spectrogram-based \mbox{X-UMX}~\cite{sawata2021all} and the waveform-based TasNet~\cite{luo2018tasnet,lancaster2020frugal} for low-latency MSS. When adapting these models for real-time processing, we observe artifacts and a considerable drop in separation quality. Therefore, we propose a novel hybrid architecture called the Hybrid Spectrogram Time-domain Audio Separation Network (HS-TasNet) that outperforms both models and demonstrates the potential for efficient real-time demixing. We perform objective and subjective evaluations to compare this real-time framework to models designed for offline processing and obtain competitive results on the MusDB dataset.

\section{State-of-the-art Architectures}
Many state-of-the-art models adopt a U-Net-style architecture~\cite{defossez2019music, jansson2017singing, kong2021decoupling, rouard2023hybrid}, where the network encodes a large context of input audio into a latent space representation, and subsequently uses a decoder to reconstruct the audio. 
In addition, architectures such as Wave-U-Net~\cite{stoller2018wave} and Demucs~\cite{defossez2019music} utilise past and future contexts of a given mixture to improve separation performance at the edges. The input receptive field for such U-Nets is in the range of 2~s to 30~s~\cite{stoller2018wave,defossez2021hybrid}, which makes it challenging to adapt to low-latency scenarios. 

The Open-Unmix~(UMX) model~\cite{stoter2019open} is a simple bidirectional LSTM-based framework that obtained state-of-the-art results on the MusDB dataset in 2018. It only adopts stacked LSTM layers and hence, does not depend on a latent space representation like U-Nets. CrossNet-Open-Unmix (\mbox{X-UMX})~\cite{sawata2021all} improved the UMX by bridging these independent networks and proposing a novel combinational and multi-domain loss. In order to develop a real-time implementation of such a model, we can replace the bidirectional LSTM with a causal unidirectional LSTM. 

The time-domain audio separation network (TasNet)~\cite{luo2018tasnet} designed for real-time speech separation in the waveform domain obtained state-of-the-art results and surpassed spectrogram-based approaches for causal and non-causal implementations. The model adopts an encoder-masker-decoder structure, where the encoder is a learned 1D convolution that creates a non-negative representation of the input. The advantage of TasNet is that it is not limited by the spectral resolution of the STFT and hence, obtains a latency as low as 5~ms. The non-causal TasNet was adopted by \cite{lancaster2020frugal} for MSS.

In Conv-TasNet~\cite{luo2019conv}, the LSTM layers are replaced by temporal convolutions. This significantly speeds up the training speed and allows the algorithm to have lower latency due to potentially smaller kernel sizes~\cite{luo2019conv}. Multiple studies have adopted the \mbox{Conv-TasNet} for MSS~\cite{defossez2019music,samuel2020meta} showing promising results. However, for real-time separation, the disadvantage of \mbox{Conv-TasNet} is that the performance of the algorithm is sensitive to the input receptive field. As explained by \cite{luo2019conv}, for 2-speaker separation, the performance of the algorithm dropped when the receptive field was reduced from 1.5~s to 0.5~s. Although a large receptive field does not reduce the algorithmic latency, it hinders computational efficiency because a larger audio block needs to be processed at every time step.

\section{Proposed Models}
\subsection{Low-Latency Constraints}
There are multiple considerations that need to be made when developing a real-time demixing system. The first factor is algorithmic latency, which is the amount of context required to output one audio frame. This could be caused by overlap-add~\cite{wang2022stft} or look-ahead windows~\cite{hu2020dccrn}. The second factor is computational efficiency, which is the time taken to process a frame of audio. For instance, if the algorithm uses too many look-behind windows or the network has many blocks that need to be sequentially processed, then the algorithm is said to have poor computational efficiency.

To ensure feasibility during real-time deployment we make certain considerations when designing models. For a given hop size, the average inference time on 4 CPU cores should be less than 50\% of the hop size, to allow reasonable room for updating audio blocks. Similar real-time constraints were imposed by the Deep Noise Suppression Challenge at ICASSP 2023~\cite{dubey2023icassp}. We are aware that this may change with increasing computational power in future research, which is beyond the scope of this study. More details on the hardware used can be found in section~\ref{sec:inference-times}.

\subsection{Low-Latency Adaptations}
\noindent {\bf X-UMX}~\cite{sawata2021all} is a spectrogram-based model whose output goes through a Wiener filter~\cite{liutkus2019norbert}, to improve separation quality and minimise interference between sources. For our causal implementation, we changed the bidirectional LSTMs to unidirectional LSTMs, reduced the window size from 4096 to 1024, and the hop size from 1024 to 512. We were unable to use the Wiener filter for real-time processing because it adopts an expectation maximization algorithm, which requires look-ahead windows for good performance~\cite{liutkus2019norbert}.

\noindent {\bf TasNet} in \cite{lancaster2020frugal} used a window size of 220~samples~(5~ms at 44.1kHz sample rate). Although studies state that smaller window sizes lead to better performance in the TasNet~\cite{luo2018tasnet, luo2019conv}, we did not observe the same trend in this causal setup. Initial experiments of hyperparameter tuning showed that a window size of 1024 performed better than smaller sizes. This is expected because the neural network has access to more context of audio at each time step. Therefore, we adopted window and hop sizes of 1024 and 512 respectively. Similar to \cite{luo2018tasnet, luo2018real}, we set the number of hidden units in the unidirectional LSTMs to 1000 and added an identity skip connection between every two LSTM layers. The number of basis signals $N$ in the convolutional layer was set to 1500. 

The final layer in the TasNet is a transposed convolution comprising learned filters. To avoid discontinuities in the audio output, we applied a Hanning window to the filters when performing the transposed convolution. This leads to a seamless overlap-add mechanism during audio reconstruction. Such an approach has also been adopted by studies that use analytic filters~\cite{pariente2020filterbank} and sinc filters~\cite{ravanelli2018speaker}. 
 
\subsection{Hybrid Spectrogram TasNet (HS-TasNet)}
Our proposed model HS-TasNet is inspired by the hybrid Demucs architecture~\cite{defossez2021hybrid} that comprises a temporal branch, a spectral branch, and shared layers. Waveform-based models are well suited to \emph{drums} and \emph{bass}, while spectrogram-based models can be better for \emph{vocals} and \emph{other}~\cite{defossez2019music}. The hybrid framework harnesses the advantages of both the spectrogram and waveform domains.

\begin{figure}[tb]
\centering
\includegraphics[width=6.5cm]{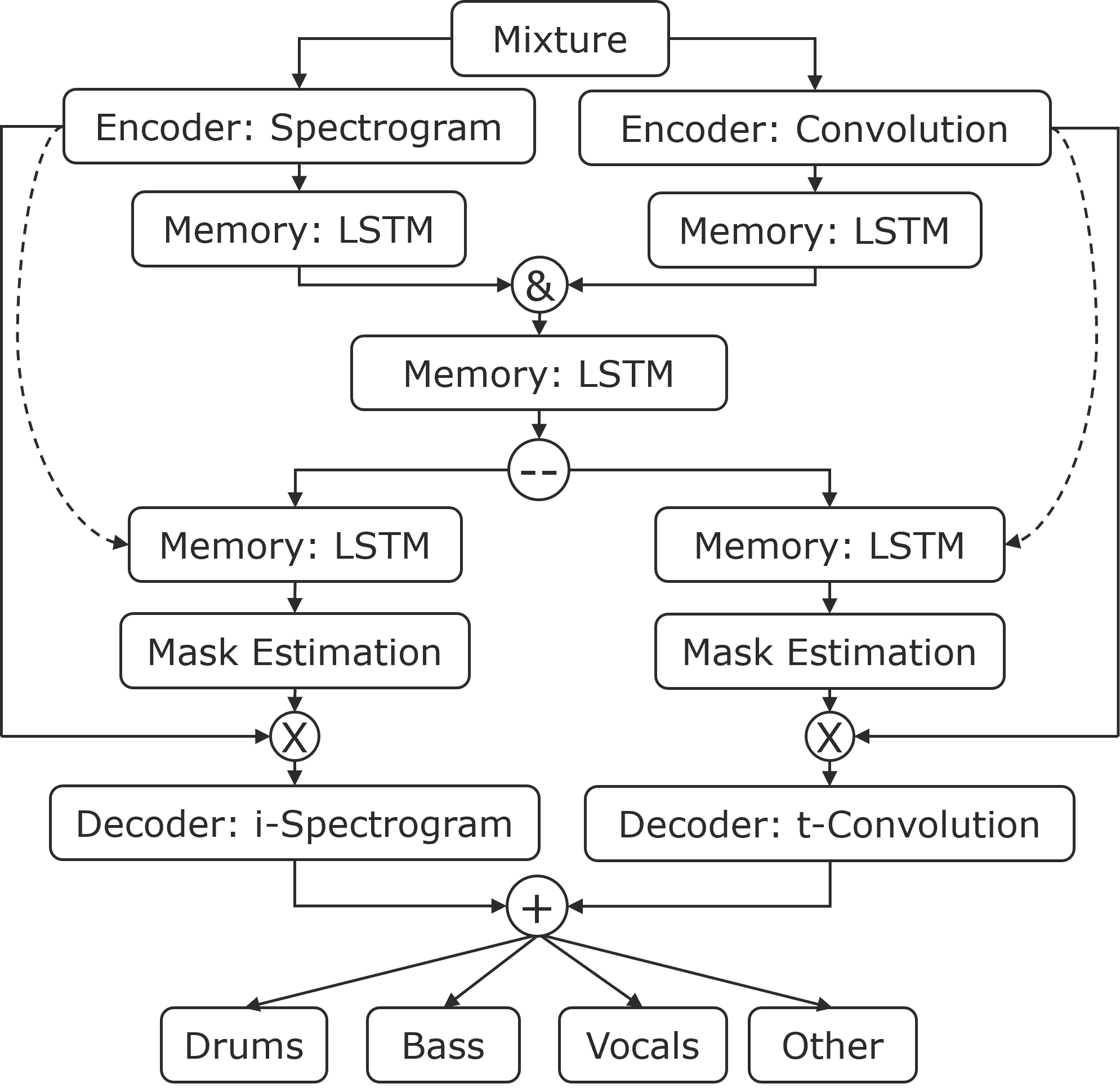}
\caption{Structure of the HS-TasNet. `\&', `-{}-', `+', `i-', `t-', and dotted lines stand for concatenate, split, sum, inverse, transpose and skip connections respectively.}
\label{fig:structure}
\vspace{-.2cm}
\end{figure}

Figure \ref{fig:structure} illustrates the architecture of the HS-TasNet. The spectrogram encoder generates 513 features as the window size is 1024. The learned convolution encoder generates 1024 basis signals. Each `Memory: LSTM' block in figure \ref{fig:structure} consists of two LSTMs with an identity skip connection added to the output of the second LSTM. The number of hidden units for the LSTMs in the spectrogram and learned convolution branches is 500. The combined branch has 1000 hidden units because the features from the individual branches are concatenated. In the `Memory: LSTM' block after the `split' layer, the encoded representation is added as a skip connection, instead of an identity one. Linear fully-connected layers were used to address any dimension mismatch between layers.

We also propose a computationally cheaper version which we refer to as \emph{HS-TasNet-Small}. Here, we replace the `concatenate' layer with a `summation'. This way, the number of the LSTM units in the combined branch is only 500, instead of 1000. Moreover, in each `Memory: LSTM' block, we use one LSTM instead of two. This reduces the number of parameters in the model from 42~M to 16~M. 

\section{Experimental Setup}
\noindent {\bf Dataset:} This study adopts the MusDB-HQ dataset~\cite{rafii2019musdbhq} comprising 86 tracks for training, 14 tracks for validation, and 50 tracks for testing. 
All audio files are stereo at 44.1~kHz sampling rate. We also trained the models using an internal dataset of 150 songs.
 
\noindent {\bf Training Strategy:} The implementation of \mbox{X-UMX} to train on MusDB is openly available in this repository\footnote{\href{https://github.com/asteroid-team/asteroid}{https://github.com/asteroid-team/asteroid}}. We used the same training pipeline and hyperparameters as the original implementation, except for changing the window size, hop size, and LSTM from bidirectional to unidirectional. Please refer to \cite{sawata2021all,sawata2023whole} for more details on the implementation. For the TasNet model, we adopted the training pipeline by DemucsV2~\cite{defossez2019music}, which is openly available in this repository\footnote{\href{https://github.com/facebookresearch/demucs/tree/v2}{https://github.com/facebookresearch/demucs/tree/v2}}, as it was more optimised to train waveform-based models. It was optimised using the L1 loss with data augmentations --- channel swapping, random gain, shuffling sources, and pitch/tempo shifting~\cite{stoter2019open, defossez2019music}. The initial learning rate was set to \num{3e-4} and was decayed by 0.5 if the validation loss did not improve for 3 epochs. Training was stopped if the validation loss did not improve for 10 epochs. We used the same set of parameters to train the HS-TasNet.

\noindent {\bf Evaluation:} To compare our models with the state-of-the-art, we use signal-to-distortion ratio~(SDR), which is popularly adopted in the literature~\cite{defossez2019music, stoller2018wave, rouard2023hybrid}. The models trained with additional data were also subjectively evaluated through listening tests, using the Multiple Stimuli with Hidden Reference and Anchor (MUSHRA) framework~\cite{schoeffler2018webmushra}. 16 participants\footnote{Participants were from L-Acoustics and an informed consent was obtained before the test.} were asked to rate 8~pages of 7~examples each (5 models, a reference and an anchor) --- 4~pages were focused on \emph{quality} (timbral fidelity and absence of artefacts) and 4~pages were focused on the rejection of \emph{interference} from remaining sources. We adapted webMUSHRA~\cite{schoeffler2018webmushra} to randomly sample examples from the MusDB test set. The \emph{quality} anchor was a distorted version of the reference, using the algorithm presented in \cite{ward2018bss}. The \emph{interference} anchor was the original mixture of audio. ANOVA and post-hoc Tukey tests were used to conduct statistical analysis. 


\section{Results}
\subsection{Evaluation}

\noindent {\bf X-UMX:} As presented in table \ref{table:result-xumx}, there is a noticeable drop in the SDR when we exclude Wiener filtering and change the LSTM from bidirectional to unidirectional. For a real-time demixer with a latency of 93~ms, the overall SDR is 4.57 which is significantly lower than the original 5.79. For a latency of 23~ms, it further drops to 3.93. This demonstrates the importance of future context and higher window sizes in MSS, which is unavailable during real-time processing. 

\begin{table}[]
\centering
\footnotesize
\vspace{-.22cm}
\caption{The X-UMX model evaluated on MusDB with and without Wiener filtering (WF). The window size (win.) of the STFT is equal to the latency of the algorithm. Uni. denotes whether the LSTM is unidirectional and RT specifies if the algorithm can run in real-time.}
\label{table:result-xumx}
\begin{tabular}{@{}ccccccccc@{}}
\toprule
WF?    & Uni?       & RT?        & \begin{tabular}[c]{@{}c@{}}Win.\\ (ms)\end{tabular} & All                  & Voc.               & Dru.                & Bass                 & Oth.                \\ \midrule
$$\cmark$$ & $$\xmark$$ & $$\xmark$$ & 93                                                  & 5.79                 & 6.61                 & 6.47                 & 5.43                 & 4.64                 \\
$$\xmark$$ & $$\xmark$$ & $$\xmark$$ & 93                                                  & 5.13                 & 6.02                 & 5.62                 & 4.58                 & 4.29                 \\ \midrule
$$\cmark$$ & $$\cmark$$ & $$\xmark$$ & 93                                                  &      5.05                &     5.73               &   5.53                 &      4.86              &  4.09                  \\
$$\xmark$$ & $$\cmark$$ & $$\cmark$$ & 93                                                  & 4.57 & 5.49 & 4.69 & 4.13 & 3.96 \\ \midrule
$$\cmark$$ & $$\cmark$$ & $$\xmark$$ & 23                                                  & 4.08                 & 4.87                 & 4.66                 & 3.76                 & 3.03                 \\
$$\xmark$$ & $$\cmark$$ & $$\cmark$$ & 23                                                  & 3.93                 & 4.65                 & 4.36                 & 3.79                 & 2.92                 \\ \bottomrule
\end{tabular}
\end{table}

\begin{figure}[t]
\centering
\includegraphics[width=8.0cm]{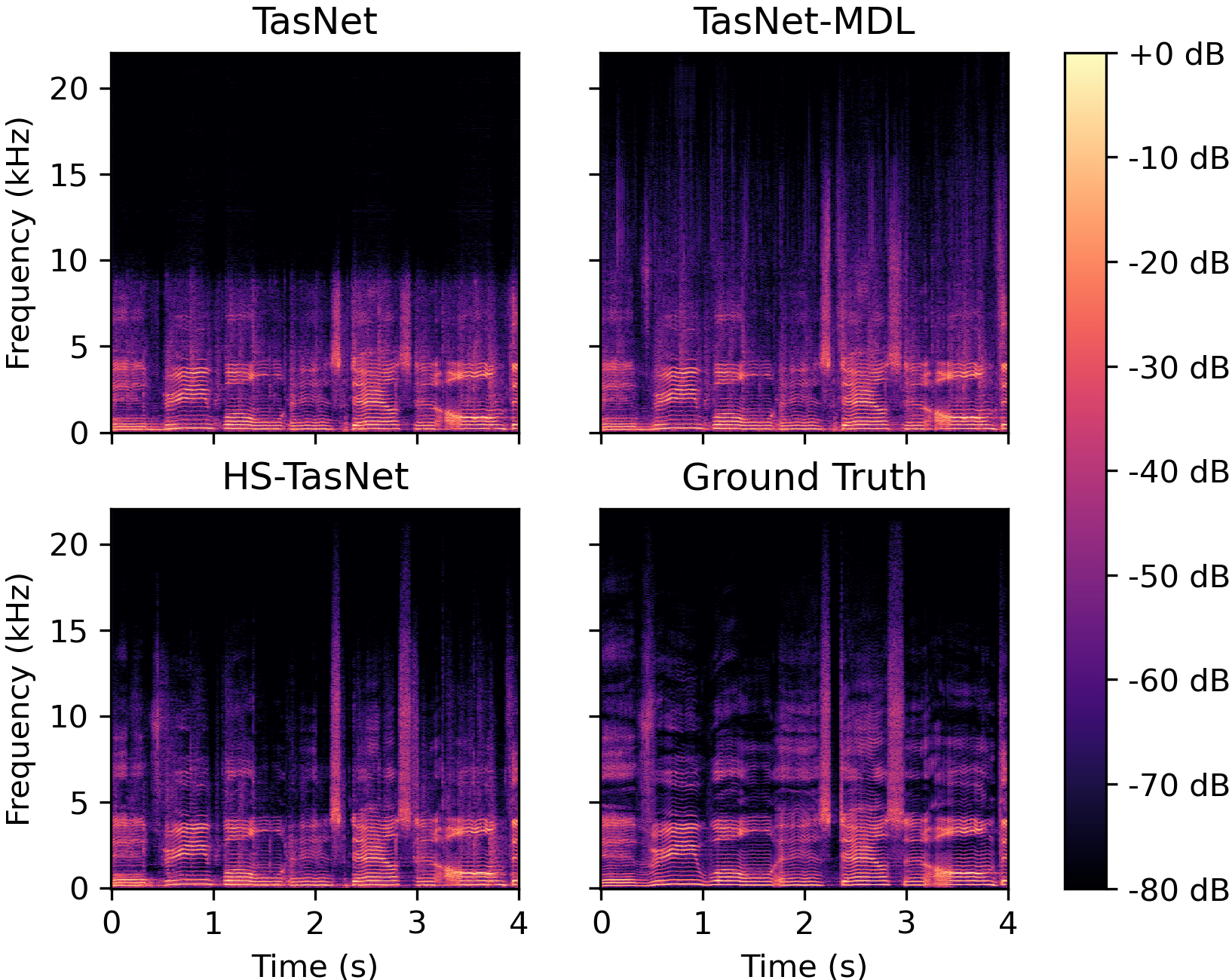}
\vspace{-.3cm}
\caption{Comparison of spectra produced by TasNet, TasNet with multi-domain loss, HS-TasNet, and the Ground Truth.}
\label{fig:hf-missing}
\vspace{-.2cm}
\end{figure}

\noindent {\bf TasNet:} As presented in Table \ref{table:result-main}, TasNet obtained a better SDR than the \mbox{X-UMX} for a latency of 23~ms. However, as shown in Figure \ref{fig:hf-missing}, we observed that the model was unable to effectively generate higher frequencies in the spectrum. Interestingly, we found that this is associated with the loss function because when we use L1 directly in the waveform domain, there is a risk of overweighting the importance of lower frequencies~\cite{moore2012introduction,wright2020perceptual}. 
Studies for speech separation have demonstrated the superior performance of scale-invariant source-to-noise ratio (SI-SNR) and scale-dependent SDR over L1/L2 for TasNets~\cite{luo2018real, heitkaemper2020demystifying}. However, in our experiments with MSS, SI-SNR and SD-SDR lead to poorer performance and slower convergence. 
We believe this might be due to the data augmentation that we have employed (which was not performed in \cite{lancaster2020frugal} during training with SI-SNR) or potentially higher correlation between sources (\cite{luo2018real} found that MSE was better than SI-SNR for dereverberation due to higher correlation between sources). We also explored the multi-domain loss~\cite{sawata2021all}, which calculates the loss on frequency and time domains. As shown in Figure \ref{fig:hf-missing}, this enabled the network to generate higher frequencies, however, the output had higher leakage between stems and did not lead to an overall improvement from using L1.

\begin{table*}[]
\centering
\footnotesize
\vspace{-.22cm}
\caption{Comparing our low-latency implementations of X-UMX, TasNet, and HS-TasNet to the state-of-the-art models. `Extra' stands for how much training data in addition to MusDB is used. 1-core, 4-core, and GPU mention the time taken by the network to process 23~ms of audio using the respective hardware averaged across 1000 iterations. ${\dagger}$ indicates from current work.}\vspace{-.15cm}
\label{table:result-main}
\begin{tabular}{@{}ccccccccccccc@{}}
\toprule
                                                                                                           & Architecture                         & All           & Vocals        & Drums          & Bass           & Other         & Extra?       & \begin{tabular}[c]{@{}c@{}}Latency\\ (ms)\end{tabular} & Param. & \begin{tabular}[c]{@{}c@{}}1-core\\ (ms)\end{tabular} & \begin{tabular}[c]{@{}c@{}}4-core\\ (ms)\end{tabular} & \begin{tabular}[c]{@{}c@{}}GPU\\ (ms)\end{tabular} \\ \midrule
\multirow{4}{*}{Non-causal models}                                                                         & Conv-TasNet~\cite{defossez2019music} & 5.73          & 6.43          & 6.02           & 6.20           & 4.27          & $$ \xmark $$ & 2,000                                                  & 9 M    & \textbf{52.29}                                        & \textbf{46.50}                                        & \textbf{16.77}                                     \\
                                                                                                           & X-UMX~\cite{sawata2021all}           & 5.79          & 6.61          & 6.47           & 5.43           & 4.64          & $$ \xmark $$ & 8,000                                                  & 36 M   & -                                                     & -                                                     & -                                                  \\
                                                                                                           & DemucsV2~\cite{defossez2019music}    & 6.28          & 6.84          & 6.86           & 7.01           & 4.42          & $$ \xmark $$ & 8,000                                                  & 288 M  & -                                                     & -                                                     & -                                                  \\
                                                                                                           & DemucsV4~\cite{rouard2023hybrid}     & \textbf{7.52} & \textbf{7.93} & \textbf{7.94}  & \textbf{8.48}  & \textbf{5.72} & $$ \xmark $$ & 7,800                                                  & 41 M   & -                                                     & -                                                     & -                                                  \\ \midrule
\multirow{5}{*}{\begin{tabular}[c]{@{}c@{}}Non-causal models \\ with extra data\end{tabular}}              & Conv-TasNet~\cite{defossez2019music} & 6.32          & 6.74          & 7.11           & 7.00           & 4.44          & 150          & 2,000                                                  & 9 M    & 52.29                                                 & 46.50                                                 & 16.77                                              \\
                                                                                                           & X-UMX~\cite{sawata2023whole}         & 6.52          & 7.57          & 7.39           & 6.28           & 4.83          & 1505         & 8,000                                                  & 36 M   & -                                                     & -                                                     & -                                                  \\
                                                                                                           & TasNet~\cite{lancaster2020frugal}    & 6.52          & 7.34          & 7.68           & 7.04           & 4.04          & 2500         & 15,000                                                 & 29 M   & \textbf{11.75}                                        & \textbf{4.20}                                         & \textbf{2.38}                                      \\
                                                                                                           & DemucsV2~\cite{defossez2019music}    & 6.79          & 7.29          & 7.58           & 7.60           & 4.69          & 150          & 8,000                                                  & 288 M  & -                                                     & -                                                     & -                                                  \\
                                                                                                           & DemucsV4~\cite{rouard2023hybrid}     & \textbf{9.20} & \textbf{9.37} & \textbf{10.83} & \textbf{10.47} & \textbf{6.41} & 800          & 7,800                                                  & 41 M   & -                                                     & -                                                     & -                                                  \\ \midrule
\multirow{4}{*}{\begin{tabular}[c]{@{}c@{}}Real-time \\ Low-Latency models\end{tabular}}                   & X-UMX$^{\dagger}$                    & 3.93          & 4.65          & 4.36           & 3.79           & 2.92          & $$ \xmark $$ & 23                                                     & 31 M   & 9.85                                                  & 4.06                                                  & \textbf{1.80}                                      \\
                                                                                                           & TasNet$^{\dagger}$                   & 4.40          & 5.02          & 4.38           & \textbf{4.73}  & 3.48          & $$ \xmark $$ & 23                                                     & 51 M   & 9.81                                                  & 4.45                                                  & 1.90                                               \\
                                                                                                           & HS-TasNet$^{\dagger}$                & \textbf{4.65} & 5.13          & \textbf{5.22}  & 4.59           & \textbf{3.64} & $$ \xmark $$ & 23                                                     & 42 M   & 9.10                                                  & 4.26                                                  & 3.90                                               \\
                                                                                                           & HS-TasNet-Small$^{\dagger}$          & 4.48          & \textbf{5.21} & 4.89           & 4.42           & 3.42          & $$ \xmark $$ & 23                                                     & 16 M   & \textbf{3.98}                                         & \textbf{1.83}                                         & 2.10                                               \\ \midrule
\multirow{4}{*}{\begin{tabular}[c]{@{}c@{}}Real-time \\ Low-Latency models\\ with extra data\end{tabular}} & X-UMX$^{\dagger}$                    & 4.10          & 4.74          & 4.62           & 3.76           & 3.28          & 150          & 23                                                     & 31 M   & 9.85                                                  & 4.06                                                  & \textbf{1.80}                                      \\
                                                                                                           & TasNet$^{\dagger}$                   & 4.92          & 5.22          & 5.54           & 5.13           & 3.78          & 150          & 23                                                     & 51 M   & 9.81                                                  & 4.45                                                  & 1.90                                               \\
                                                                                                           & HS-TasNet$^{\dagger}$                & \textbf{5.55} & \textbf{5.97} & \textbf{6.34}  & \textbf{5.62}  & \textbf{4.28} & 150          & 23                                                     & 42 M   & 9.10                                                  & 4.26                                                  & 3.90                                               \\
                                                                                                           & HS-TasNet-Small$^{\dagger}$          & 5.01          & 5.51          & 5.75           & 4.93           & 3.86          & 150          & 23                                                     & 16 M   & \textbf{3.98}                                         & \textbf{1.83}                                         & 2.10                                               \\ \bottomrule
\end{tabular}
\vspace{-.35cm}
\end{table*}

\noindent {\bf HS-TasNet} obtained an overall SDR of 4.65 and surpassed the TasNet's performance for all four sources. As shown in Figure \ref{fig:hf-missing}, there were also no issues with generating higher frequencies. The smaller version of the HS-TasNet obtains an overall SDR of 4.44. 
In Table \ref{table:result-main}, we present the SDR scores of X-UMX, TasNet, and HS-TasNet compared to the state-of-the-art models. The HS-TasNet scales better to additional training data. The overall SDR is 5.55 when trained on additional data, versus 4.92 for the TasNet and 4.10 for the \mbox{X-UMX}. The HS-TasNet-Small obtains an overall SDR of 5.01, which is still higher than TasNet and X-UMX.

\noindent {\bf Listening test} results in table~\ref{table:result-mushra} show that DemucsV4 significantly outperforms other models for \emph{quality} and \emph{interference} ($p < 0.001$). \mbox{HS-TasNet} was significantly better than TasNet~($p < 0.005$). There were no significant differences between \mbox{HS-TasNet}, DemucsV2, and \mbox{Conv-TasNet} ($p > 0.1$). For \emph{quality}, the \mbox{HS-TasNet} is slightly behind DemucsV2 and equal to \mbox{Conv-TasNet}. \mbox{HS-TasNet} surpasses DemucsV2 and \mbox{Conv-TasNet} for the rejection of \emph{interference} from remaining sources. Audio examples are available here\footnote{\href{https://l-acoustics.github.io/demix.github.io/}{https://l-acoustics.github.io/demix.github.io/}}.

\subsection{Inference Times}
\label{sec:inference-times}
In Table~\ref{table:result-main}, we also present the time taken by the algorithm to run on an Intel i7-12850HX CPU on 1-core and 4-cores, and NVIDIA RTX 3080Ti GPU. The offline versions of TasNet~\cite{lancaster2020frugal} and Conv-TasNet~\cite{defossez2019music} cannot run in real-time, but we still present the time taken to process 1024 samples. Although the Conv-TasNet has only 9~M parameters, it has many blocks that require sequential processing, which makes it computationally inefficient. The other offline models like Demucs cannot process a block as small as 1024 due to the large number of layers. 
The inference time of all our real-time models is around 4~ms when running on 4-cores, which gives it enough time to update within the hop size of 11~ms. The HS-TasNet-Small can run comfortably even on a single CPU core. 



\begin{table}[t]
\centering
\footnotesize
\vspace{-.2cm}
\caption{The mean ($\pm$ standard deviation) from the listening test. The upper half is for \emph{quality} and the bottom half is for rejection of \emph{interference}. The bold values indicate the highest value excluding the reference and DemucsV4. ${\dagger}$ indicates from current work.
}
\label{table:result-mushra}
\begin{tabular}{@{}cccccc@{}}
\toprule
Model                                & All                                & Vocals                             & Drums                              & Bass                               & Other                              \\ \midrule
Reference                            & 99\textsubscript{$\pm$18}          & 99\textsubscript{$\pm$2}           & 95\textsubscript{$\pm$18}          & 100\textsubscript{$\pm$0}          & 100\textsubscript{$\pm$0}          \\
DemucsV4~\cite{rouard2023hybrid}     & 66\textsubscript{$\pm$26} & 70\textsubscript{$\pm$23} & 67\textsubscript{$\pm$24} & 49\textsubscript{$\pm$29} & 78\textsubscript{$\pm$26} \\
DemucsV2~\cite{defossez2019music}    & \textbf{39\textsubscript{$\pm$23}}          & 34\textsubscript{$\pm$22}          & \textbf{48\textsubscript{$\pm$27}}          & \textbf{34\textsubscript{$\pm$24}}          & 40\textsubscript{$\pm$23}          \\
HS-TasNet$^{\dagger}$               & 37\textsubscript{$\pm$23}          & \textbf{44\textsubscript{$\pm$23}}          & 39\textsubscript{$\pm$21}          & 29\textsubscript{$\pm$21}          & 38\textsubscript{$\pm$23}          \\
Conv-TasNet~\cite{defossez2019music} & 37\textsubscript{$\pm$31}          & 39\textsubscript{$\pm$21}          & 44\textsubscript{$\pm$26}          & 21\textsubscript{$\pm$23}          & \textbf{43\textsubscript{$\pm$31}}          \\
TasNet$^{\dagger}$                 & 27\textsubscript{$\pm$23}          & 26\textsubscript{$\pm$13}          & 27\textsubscript{$\pm$18}          & 25\textsubscript{$\pm$25}          & 30\textsubscript{$\pm$23}          \\
Anchor~(Quality)                   & 5\textsubscript{$\pm$9}            & 4\textsubscript{$\pm$8}            & 1\textsubscript{$\pm$3}            & 10\textsubscript{$\pm$16}          & 5\textsubscript{$\pm$9}            \\ \midrule
Reference                            & 99\textsubscript{$\pm$4}           & 99\textsubscript{$\pm$5}           & 100\textsubscript{$\pm$0}          & 100\textsubscript{$\pm$1}          & 99\textsubscript{$\pm$4}           \\
DemucsV4~\cite{rouard2023hybrid}     & 78\textsubscript{$\pm$12} & 79\textsubscript{$\pm$15} & 86\textsubscript{$\pm$12} & 64\textsubscript{$\pm$29} & 83\textsubscript{$\pm$12} \\
DemucsV2~\cite{defossez2019music}    & 52\textsubscript{$\pm$22}          & 42\textsubscript{$\pm$20}          & 57\textsubscript{$\pm$18}          & 61\textsubscript{$\pm$21}          & 49\textsubscript{$\pm$22}          \\
HS-TasNet$^{\dagger}$              & \textbf{58\textsubscript{$\pm$23}}          & \textbf{54\textsubscript{$\pm$22}}          & \textbf{60\textsubscript{$\pm$20}}          & \textbf{62\textsubscript{$\pm$23}}          & 55\textsubscript{$\pm$23}          \\
Conv-TasNet~\cite{defossez2019music} & 48\textsubscript{$\pm$26}          & 36\textsubscript{$\pm$18}          & 53\textsubscript{$\pm$25}          & 46\textsubscript{$\pm$22}          & \textbf{56\textsubscript{$\pm$26}}          \\
TasNet$^{\dagger}$                 & 47\textsubscript{$\pm$22}          & 40\textsubscript{$\pm$19}          & 45\textsubscript{$\pm$20}          & 61\textsubscript{$\pm$25}          & 43\textsubscript{$\pm$22}          \\
Anchor~(Interference)              & 3\textsubscript{$\pm$25}           & 0\textsubscript{$\pm$1}            & 1\textsubscript{$\pm$3}            & 2\textsubscript{$\pm$5}            & 9\textsubscript{$\pm$25}           \\ \bottomrule
\end{tabular}
\vspace{-.2cm}
\end{table}

\section{Conclusion}
This paper explored real-time low-latency demixing for vocals, drums, bass, and other. A few challenges associated with this task include the non-availability of future context and the poor computational efficiency of U-Nets. Initially, we adapted the X-UMX and TasNet models for low-latency demixing. We observed that the performance of the spectrogram-based model drops with reducing the window size due to the spectral resolution of the STFT. Although the TasNet performed better than the X-UMX, we observed artifacts such as missing higher frequencies. Therefore, we proposed a novel model called the HS-TasNet that harnesses the advantages of both spectral and waveform domains. It obtained an overall SDR of 4.65 when trained only on MusDB and increased to 5.55 with additional training data. Listening tests showed that the HS-TasNet significantly outperformed the TasNet and it was comparable to DemucsV2, which was the state-of-the-art model two years ago~\cite{defossez2019music}. These results demonstrate the potential of efficient demixing for real-time low-latency music applications.

This study considered 23~ms to be the required latency for the demixing models. Future work can evaluate the feasibility of further reducing the latency, which would make it applicable to a variety of live music scenarios. 
Moreover, training strategies such as decomposing/reorganizing a Bi-RNN layer~\cite{li2023design} and using non-causal models to train causal models~\cite{drossos2018mad} can mitigate the performance degradation when changing non-causal models to their causal counterparts.

\section{Acknowledgments}
The authors thank Emmanouil Benetos from  Queen Mary University of London and Selim Sheta, Thomas Fouchard, Nicolas Epain, Etienne Corteel, and Guillaume Le Nost from L-Acoustics for their feedback and fruitful discussions.

\vfill\pagebreak

\footnotesize
\bibliographystyle{IEEE}
\bibliography{refs}

\end{document}